\documentstyle[12pt]{article}

\def\Real{{\bf R}}
\def\Re{{\rm Re}}
\def\Im{{\rm Im}}
\def\RE{\Re}
\def\IM{\Im}
\def\r{{\cal R}}

\def\g{{\bf g}}
\def\h{{\bf h}}
\def\vol{{\rm vol}}
\def\he{\hat{e}}
\def\hxi{\hat{\xi}}
\def\hstar{\hat{\star}}
\def\hD{\hat{D}}
\def\hphi{\hat{\phi}}
\def\homega{\hat{\omega}}
\def\hR{{\hat{R}}}
\def\tR{{\tilde{R}}}
\def\hOmega{{\hat \Omega}}
\def\hJ{{\hat{J}}}
\def\be {\begin{equation}}
\def\ee {\end{equation}}
\def\ba{\begin{eqnarray}}
\def\ea{\end{eqnarray}}
\def\bd{\begin{displaymath}}
\def\ed{\end{displaymath}}
\def\R{{\bf R}}
\def\te{{\tilde e}}
\def\tphi{{\tilde \phi}}
\newcommand{\bs}{}

\def\Nom{\homega_{CNC}}

\begin{document}
\title{The Normal Conformal Cartan Connection and the Bach Tensor}
\author{Miko\l{}aj\ Korzy\'nski${}^{1}$  and Jerzy\ Lewandowski${}^{2,1,3}$}

\maketitle
{\it 1. Instytut  Fizyki Teoretycznej, Uniwersytet
Warszawski, ul. Ho\.{z}a 69, 00-681 Warsaw, Poland\\
2. Center for Gravitational Physics and Geometry, Physics
Department, 104 Davey, Penn State, University
Park, PA 16802, USA\\
3. Max--Planck--Institut f\"ur Gravitationsphysik, Albert--Einstein--Institut,
 14476 Golm, Germany}


\begin{abstract}
The goal of this paper is to express the Bach tensor of a four--dimensional 
conformal geometry of 
an arbitrary signature by the Cartan normal conformal (CNC) connection. 
We show that the Bach tensor can be identified with the Yang--Mills current 
of the connection. It follows from that result that a conformal 
geometry whose CNC connection is reducible
in an appropriate way has a degenerate Bach tensor. As an example
we study the case of a CNC connection which admits a  twisting 
covariantly constant twistor field. This class of conformal geometries 
of this property is known as given by  the Fefferman  metric tensors.
We use our result to calculate the Bach tensor of an arbitrary Fefferman metric
and show it is proportional to the tensorial square of the four--fold 
eigenvector of the Weyl tensor. Finally, we solve the Yang--Mills equations
imposed on the CNC connection for all the homogeneous Fefferman metrics.
The only solution is the Nurowski-Pleba\'nski metric.   
\end{abstract}
\medskip
\indent\emph{Pacs:} 04.20.-q, 02.40.Ky

\section{Introduction}
In the general relativity,  conformally invariant features of a
metric are usually described by the Weyl  and the Bach tensors,
and in 3 dimensions (where the Weyl tensor identically vanishes)
by the Cotton--York tensor. Systematic approach to the
conformal geometry theory on the other hand leads to the definition of so called
\textit{Cartan normal conformal connection} \cite{kob}. In that framework
the Weyl tensor is a part of the curvature of the connection, and
in 3 dimensions the curvature consists of the Cotton--York tensor. 
An explicit relation between the Bach tensor and the connection,
however, seems not to be known in the literature. A good
suggestion follows from Merkulov's result reported in
\cite{merkulov}. The paper contains a theorem saying that the
Yang--Mills equations imposed on the twistor connection amount to
the condition that the Bach tensor of the corresponding metric
tensor is identically zero. For the proof and calculation, however,
Merkulov  referred   to
his unpublished work. 

In this paper we are not going to consider the
twistor connection \emph{per se}, hence we only recall  that
Penrose uses the Cartan normal conformal connection  to define the
covariant derivative of so called local twistor fields \cite{rp,f}
in 4--real dimensional spacetime of the Lorenzian signature. The
resulting connection is called the twistor connection.  

The first
goal of the current work is expressing the Bach tensor -- vanishing
or not -- by the Cartan normal conformal connection in an explicit
way.  We introduce here the Yang--Mills current of the Cartan
normal conformal connection of a 4--dimensional conformal geometry
of {\it arbitrary signature}, calculate the current and show it
can be identified with the Bach tensor. In particular, it follows
from our result that the Bach tensor of a conformal geometry
whose Cartan normal conformal connection is reducible is
degenerate for appropriate types of the reducibility. The best
known example of the reducibility \cite{lionel} is the class of
the conformally Einstein geometries; the Bach tensor is
identically zero in this case. Conformal
geometry which admits covariantly constant local twistor field is another
example.
The following features can be used as an equivalent definition
 of such geometries \cite{rp,lewand1}:
1) the existence of a null Killing vector field, say $k$, and 2)
the Petrov type N of the Weyl tensor. All the
4--dimensional Lorenzian geometries of this property are known
\cite{lewand1}. When the four--fold eigenvector $k$ of the Weyl tensor is
twisting, the metrics set the class of the Fefferman  metrics
\cite{feff,bds}. Each Fefferman metric tensor (or rather its conformal
class) is assigned to a 3--dimensional Cauchy--Riemann geometry. In this case the $so(2,4)$
Cartan normal conformal connection reduces to the
$su(1,2)$ Cartan--Chern connection of the corresponding
Cauchy--Riemann geometry \cite{bds}. The Yang--Mills current of
the Cartan--Chern connection with respect to the Fefferman metric
was derived in \cite{lewnur}. It is given by a single real
function density.
 The existence of a homogeneous Cartan--Chern solution of the
Yang--Mills equations was  mentioned in \cite{lewnur}. In Section 4
we correct two misprints which occurred in that work: a mistake
in Yang--Mills equation and in the Bianchi type of the solution.

The
Fefferman metrics are also known {\it not} to be conformally
Einstein except the conformally flat metric \cite{lewand2}.
Comparing those facts explains why the first (and only) known
explicit example of the non conformally Einstein and Bach flat
metric could be recently found by Nurowski and Pleba\'nski
\cite{nurpleb} in the Fefferman class.

 Using the Yang--Mills
current of the Cartan--Chern connection we derive below the Bach
tensor of a general Fefferman metric and show it is proportional
to $k\otimes k$.  Finally, we solve completely the Yang--Mills
equations imposed on the Cartan--Chern connection in the
homogeneous case. We find out that the only solution is the one
corresponding to the Nurowski--Pleba\'nski metric.  

The problem
of a conformally invariant condition for a metric tensor to be
conformally Einstein was addressed by several authors, including 
Kozameh, Newman and Tod
\cite{knt}. A sufficient set of conditions they formulated
consists of the Bach flatness plus a certain extra condition on
the Weyl tensor. However, their result concerns only the case of
appropriately generic Weyl tensor, whereas the Weyl tensor of the
Fefferman metric is degenerate in that  sense.

Our work was motivated by the paper of Nurowski and Pleba\'nski
\cite{nurpleb} (see also the CQG Highlights of 2000/2001). We hope
our result may be applied for example in the null surface
formulation of gravity of Newman and collaborators \cite{fkn}
which involves the conformal geometry, the Bach tensor and
recently also the Cartan normal conformal connection \cite{fknn}.

Both Cartan normal conformal connection and twistor connection
found recently yet another application to General Relativity as
they were used to formulate the conformal representation of
the Einstein equations, very useful when dealing with characteristic 
initial value problems
 \cite{f2,frauensp}.

Our paper should also encourage to classify and study all the
possible reductions of the Cartan normal conformal connections.

\section{The Cartan normal conformal  connection}
We recall in this section the definition of the Cartan normal
conformal connections \cite{kob,bds}. The connections correspond to
conformal
geometries. We consider here the 4-dimensional case of an arbitrarily
fixed
signature $(p,q)$. A short, working definition formulated
in the first subsection is just an appropriate assignment
of a certain $so(p+1,q+1)$ valued 1-form to every
co-frame cotangent to $M$.  Whenever two co-frames are related by a
conformal  transformation, the corresponding 1-forms are gauge
equivalent. This formulation is analogous to the definition
of the twistor connection \cite{rp,f}.
The full, geometric definition of \cite{bds} is recalled in the
second subsection.

We will use below a fixed real $4$ by $4$ symmetric matrix $\eta
=(\eta_{\mu\nu})_{\mu,\nu=1,...,4}$ of the signature $p$ minuses
and $q$ pluses, and the following 6 by 6 matrix
$Q=(Q_{IJ})_{IJ=0,...,5}$
\begin{displaymath}
Q\ =\ \left(\begin{array}{ccc}
0 & 0 & -1 \\
0 & \eta & 0 \\
-1 & 0 & 0\end{array}\right).
\end{displaymath}
The group $SO(q+1,p+1)$ will be identified with the group $SO(Q)$
of 6 by 6 real matrices
$A=({A^\mu}_\nu)^{\mu=1,...,4}_{\mu=1,...,4}$ such that
\be {A^\alpha}_\mu {A^\beta}_\nu Q_{\alpha\beta}\ =\ Q_{\mu\nu}.
\ee

\subsection{The short definition}
\label{2.1} Let $(\theta^1,\ldots,\theta^4)$ be
a sequence of differential 1-forms which sets a basis of $T^*_mM$ at every
point  $m$ of some open subset ${\cal U}\subset M$. We call it a co-frame
and assign to $(\theta^1,\ldots,\theta^4)$ a metric tensor $g$,
\be\label{g} g\ =\ \eta_{\mu\nu}\theta^\mu\otimes \theta^\nu. \ee
The frame of tangent vectors dual to $(\theta^1,\ldots,\theta^4)$ will be denoted by 
$(X_1,\ldots,X_4)$.

{\it The Cartan normal
conformal (CNC) connection 1-form in the natural gauge} is the following
matrix of 1-forms assigned to the co-frame $(\theta^1,...,\theta^4)$
\begin{equation}\label{omegaN}
\omega_{(C)}= \left(
\begin{array}{ccc}
0 & \eta_{\nu\alpha}\theta^\alpha & 0 \\ &&\\
 P^\mu &  \Gamma^\mu\!_\nu & \theta^\mu \\ &&\\
0 & \eta_{\nu\alpha} P^\alpha & 0
\end{array}\right) \label{eq:Nom}
\end{equation}
where each component labeled by $\mu$ ($\nu$) stands for a column (row)
given
by all the values $\mu=1,...,4$ ($\nu=1,...,4$),
  ${\Gamma^\mu}_\nu$ is the Riemann connection 1-form, i. e.
\be d\theta^\mu + {\Gamma^\mu}_\nu\wedge \theta^\nu\ =\ 0, \ \
\eta_{\mu\alpha}{\Gamma^\alpha}_\nu\ =\
-\eta_{\nu\alpha}{\Gamma^\alpha}_\mu, \ee
and $P^\mu$ is given by the Ricci tensor $R_{\mu\nu}
\theta^\mu\otimes \theta^\nu$ of the metric $g$, and by its Ricci scalar $R\
=\ {R^\mu}_\mu$,
\be P_\mu\ = \left(\frac{1}{12}\,R\,\eta_{\mu\nu} -
\frac{1}{2}\,R_{\mu\nu}\right)\theta^\nu \ee

The advantage of this definition is that  $\omega'_{(C)}$ assigned to a 
co-frame
$({\theta'}^1,...,{\theta'}^4)$ given by a point $m\in M$ depending conformal
transformation
\be\label{epr}{\theta'}^\mu = {\Lambda^\mu}_\nu \theta^\nu, \ee
is gauge equivalent to $\omega_{(C)}$, namely
\be\label{omegapr} {\omega'}_{(C)}\ =\ h^{-1}\omega_{(C)}h\ +\ h^{-1}dh \ee
where in the pure rescaling case (\emph{i.e.} when ${\Lambda^\mu}_\nu\ =\
c\,{\delta^\mu}_\nu$) the matrix $h$ is given by the function $c$
in the following way,
\be h\ =\ \left(\begin{array}{ccc}
c^{-1} & 0 & 0 \\ && \\ c^{-2}\,c,_\sigma \eta^{\mu\sigma} & \delta^\mu\!_\nu &
0 \\ && \\
\frac{1}{2c^3}\,\eta^{\alpha\beta}\,c,_\alpha\,c,_\beta & c^{-1}\,c,_{\nu}
& c\end{array}\right),\ee
and in the (pseudo)rotation case (\emph{i.e.} $g'\ =\ g$), the matrix
$h$ is
\be h\ =\
\left(\begin{array}{ccc}1 & 0 & 0 \\  0 & \Lambda^\mu\!_\nu & 0 \\
0 & 0 & 1 \end{array}\right).\ee
The curvature of the connection $\omega_{(C)}$ is
\begin{equation}
\Omega_{(C)} = d\omega_{(C)} + \omega_{(C)}\wedge\omega_{(C)}\ = \left(
\begin{array}{ccc}
0 & 0 & 0 \\ &&\\
D P^\mu &  C^\mu\!_\nu & 0 \\ &&\\
0 & D \eta_{\nu\alpha}P^\alpha & 0
\end{array}\right) \label{eq:NOM}
\end{equation}
where
\be
D P^\mu = d P^\mu + \Gamma^\mu\!_\nu\wedge P^\nu
\ee
and the tensor $X_\mu \otimes \theta^\nu \otimes C^\mu\!_\nu $ 
 is the
Weyl tensor of the metric $g$ (\ref{g}). Owing to the
transformation law (\ref{omegapr}) the curvature $\Omega'_N$ of
the connection $\omega'_N$ is \be \Omega_{(C)}\ =\ h^{-1}\Omega_{(C)} h.
\ee

The formula  (\ref{omegaN}) for $\omega_{(C)}$ shows the relation
between the CNC connection and the Penrose
twistor connection \cite{rp,f}. The normal gauge will be applied
in the next section in the derivation of the Yang--Mills current
of the CNC connection.

Finally, {\it Cartan normal conformal connection 1-form in a
general gauge} is given by imposing on $\omega_{(C)}$ in a natural
gauge, any transformation of the following form,
\be\label{gauge} \omega\ =\ h^{-1}\omega_{(C)} h + h^{-1}dh \ee
where $h$ is any matrix--valued function defined (locally) on $M$
which takes values in the subgroup $H\subset SO(Q)$ defined by the
matrices of the form $(\ref{H})$ (see the next subsection).

\subsection{The geometric definition}\label{2.2}
 The definitions presented in this section  are based on \cite{bds}.

{\bf The conformal bundle.}
 Given a conformal geometry $[g]$ on a
4-manifold $M$ we denote by $[g](m)\subset T^*_mM\otimes
T^*_mM$ the set of the conformally equivalent, non--degenerate symmetric 
tensors
corresponding to the conformal geometry $[g]$ at $m$. Consider the
principal fiber bundle $p:C\rightarrow M$ where
\be C\ =\ \bigcup_{m\in M}[g](m)\ \subset T^*M\otimes T^*M \ee
and the structure group is $\R^+$ acting by the rescaling
\begin{displaymath}
C\ni g(m) \mapsto \tR_{c}\,g(m) = c^2\cdot g(m) \in C.
\end{displaymath}

There is a natural horizontal metric tensor $\tilde g$ on $C$ and
a vertical vector field $\zeta$  defined by
\ba \tilde g(X,Y)\ &=&\ g(p_*\,X,p_*\,Y),\nonumber\\
\zeta(f)\ &=&\  \left.\frac{d}{dc}\right|_{c=1} {\tR_c}^*f, \ea
where $X,Y\in T_{g(m)}C$ and $f$ is a function defined on $C$, all
arbitrary. A co-frame $(\te^1,...,\te^4,\tphi)(g(m))$ at a point
$g(m)\in C$  will be called {\it admissible} if it satisfies each
of the following two conditions:
\begin{enumerate}
\item $\tilde g = \eta_{\mu\nu}\te^\mu\otimes\te^\nu$
\item $\tilde\phi(\zeta) = -1$.
\end{enumerate}
The admissible co-frames set up a conformal bundle $\pi: \
P(C)\rightarrow\ M$ defined as follows. $P(C)$ is the space of all
the admissible co-frames cotangent to $C$. The projection is
\be \pi \ :\ (\te^1,...,\te^4,\tphi)(g(m))\mapsto\ m. \ee
In $P(C)$ one considers the following three groups of maps
\be (\te^1,...,\te^4,\tphi)(g(m))\ \mapsto\
(\te'^1,...,\te'^4,\tphi')(g'(m)): \ee
\begin{eqnarray}
(g', \te'^\mu, \tphi')\ &=&\ (g, \te^\mu,\tphi +
\te^\mu\,b_\mu)\label{M}\\
(g', \te'^\mu, \tphi')\ &=&\
(g,\te^\nu\,(\Lambda^{-1})^\mu\!_\nu,\tphi)\label{O}\\
(g', \te'^\mu, \tphi')\ &=&\ (c^{-2}\cdot g,\ c^{-1}\cdot
R_{c}^*\,\te^\mu, R_{c}^*\,\tphi)\label{d} \\
\end{eqnarray}
where the co-vector  $(b_\mu)_{\mu=1,...,4}$, the  matrix
$({\Lambda^\mu}_\nu)^{\mu=1,...,4}_{\nu=1,...,4}$
such that $\Lambda^T\eta\Lambda\ = \eta$ and the number $c$ are arbitrary.
The maps can
be considered as given by a right  action in $P(C)$ of elements of
$SO(p,q)$ of the following form,
%
\be\label{H}   \left(\begin{array}{ccc} 1 & 0 & 0 \\ b^\mu &
\delta^\mu\!_\nu &
0 \\ \frac{1}{2}\,b_\sigma\,b^\sigma & b_\nu & 1
\end{array}
\right),\left(\begin{array}{ccc}
1 & 0 & 0 \\
0 & \Lambda^\mu\!_\nu & 0 \\
0 & 0 & 1
\end{array}\right),\ \left(
\begin{array}{ccc}
c & 0 & 0 \\
0 & \delta^\mu\!_\nu & 0 \\
0 & 0 & \frac{1}{c}
\end{array}\right) \ee respectively. In this way the subgroup $H\subset
SO(Q)$ generated by all the matrices (\ref{H}) acts on $P(C)$.
This completes the construction of the $H$-principal fiber bundle
structure  of $\pi:P(C)\rightarrow M$.

Using the following map $\lambda: P(C)\rightarrow C$,
\be
\lambda \ :\ (\te^1,...,\te^4,\tphi)(g(m))\mapsto\ g(m).
\ee
we lift to $P(C)$ the co-frames defined on $C$,
 \be \he^\mu\ =\
\lambda^*\,\te^\mu,\ \ \hphi\ =\ \lambda^*\,\tphi. \ee

\bigskip

{\bf The connection.} We denote here by $\hR$ the right action of
the group $H$ in $P(C)$. Due to $\hR$ every left invariant vector
field $\xi$ tangent to the $H$ defines naturally a vector field $\hxi$
tangent to $P(C)$. Since $H$ is the subgroup of the group of the
matrices $SO(Q)$, we identify the Lie algebra $\h$
 of the left
invariant vector fields on $H$ with the corresponding algebra of
matrices. Since ${\rm dim}\,M + {\rm dim}\,H\ =\ {\rm
dim}\,SO(Q)$, one can consider $SO(Q)$-Cartan connections on the
principal fiber bundle $(P(C), M, H, \pi)$. (A detailed
introduction to the Cartan connections  can be found in
\cite{kob}). An  $SO(Q)$-Cartan connection defined on $P(C)$ is an
$so(Q)$  valued 1-form $\homega$ defined on $P(C)$ such that the
following three conditions are satisfied:
\begin{enumerate}
\item $\homega(X) = 0 \iff X=0$ (the non-degeneracy)
\item $\homega(\hxi) = \xi$ for every $\xi\in \h$
\item $\hR_h\!^*\homega = h^{-1}\homega h$ for every
$h\in H$.
\end{enumerate}
One of the connections, say $\homega_{(C)}$, can be uniquely
distinguished by the following two natural conditions. The first
one concerns the first row of $\homega_{(C)}$, namely we require that
\begin{displaymath}
\homega_{(C)} = \left(\begin{array}{ccc}
\,* & \he^\mu\eta_{\mu\nu} & * \\ && \\
\,* & * & *\\ && \\
\,* & * & * \end{array}\right)
\end{displaymath}
(* denotes those components which are arbitrary modulo
the symmetries of the matrices of the Lie algebra $so(Q)$). The
second condition is imposed on the curvature of $\homega_{(C)}$,
namely one requires that
\begin{equation}
\hOmega_{(C)} :=d\Nom + \Nom\wedge\Nom = \left(\begin{array}{ccc} 0 & 0 & 0 \\
\,* & \frac{1}{2}\,K^\mu\!_{\nu\rho\sigma} \he^\rho\wedge\he^\sigma & 0
\\ 0 &*& 0\end{array}\right) \label{eq:myres}
\end{equation}
and
\begin{equation}
K^\mu\!_{\nu\mu\sigma} = 0. \label{eq:bysser}
\end{equation}

A  section $\sigma: {\cal U} \rightarrow P(C)$, where ${\cal
U}\subset M$, defines the pullback of $\homega_{(C)}$ onto ${\cal U}$,
\begin{displaymath}
\sigma^*(\homega_{(C)})\ =\ \left(\begin{array}{ccc}
-\psi & \theta^\alpha\eta_{\alpha\nu} & 0 \\ && \\
V^\mu & G^\mu\!_\nu &  \theta^\nu \\ && \\
0 & V^\alpha \eta_{\alpha\nu} & \psi \end{array}\right).
\end{displaymath}
The pullback is in general different than $\omega_{(C)}$ defined in
the previous subsection. However, for a 1-form $b$ defined
on ${\cal U}$, we consider at each $m\in M$
the corresponding $h_{b(m)}\in H$ given by the first matrix in
(\ref{H}) and a new section
\be
\sigma'=\hR_{h_b}^*\sigma \label{eq:sigma'}
\ee
The corresponding diagonal element $\psi'$ of $\sigma'^*\homega$ is
\bd \psi'\ =\ \psi+b. \ed
Since $b$ is arbitrary, we can fix it such that
\be \psi' =\ 0. \ee
With respect to this {\it natural section} $\sigma'=:\sigma_N$
\be\sigma_N^*\homega_{(C)}\ =\ \omega_{(C)}, \ee
that is  $\sigma_N*\homega$ is exactly the CNC
connection 1-form in the natural gauge defined in (\ref{omegaN}).

\section{The Yang--Mills current of $\homega_{(C)}$}
Given a conformal geometry $[g]$ on $M$, there is a uniquely
defined Hodge  $\star$ acting in the space of the differential
2-forms defined on $M$ \footnote{This is true for dimension $d=4$. Otherwise the conformal geometry is not
enough to determine the action of $\star$}. We use it below to define the Yang--Mills
current of the  CNC connection as the source
given by imposing the Yang--Mills equations. As in the previous
section, we formulate the definition in two versions: directly on
$M$ and, respectively, on the bundle $P(C)$. We use the first one
to calculate the Yang--Mills current. Next, we
find the formula for the Yang--Mills current defined on $P(C)$.

\subsection{The calculation in the terms of the working definition}
Given a co-frame $(\theta^1,...,\theta^4)$ and the corresponding Cartan
normal conformal connection 1-form $\omega_{(C)}$ in the natural
gauge, the {Yang--Mills current}  of $\omega_{(C)}$ is
\be J_{(C)}\ :=\ D\star\Omega_{(C)}\ =\ d\star\Omega_{(C)} + \omega_{(C)}\wedge
\star\Omega_{(C)} - \star\Omega_{(C)}\wedge\omega_{(C)}. \label{eq:current}\ee
The conformal transformations (\ref{epr},\ref{omegapr}) of the
co-frame and the connection, respectively, are accompanied by the
suitable gauge transformation of $J_{(C)}$, namely
\bd J_{(C)}\ =\ h^{-1}J_{(C)} h, \ed
where $h$ is the same is in (\ref{omegapr}).

Substituting $\Omega_{(C)}$ and $\omega_{(C)}$ in (\ref{eq:current}) for
those given by (\ref{eq:NOM}), and (\ref{eq:Nom}), respectively,
we obtain:
\begin{eqnarray}
J_{(C)}&=&\left({J_{(C)}}^K_L\right)^{K=0,...,5}_{L=0,...,5}=\nonumber\\
&=&
\left(\begin{array}{ccccc} \theta_\sigma\wedge{\star D P^\sigma} &;&
\theta_\sigma\wedge{\star
C^\sigma\!_\nu}  &;& 0 \\&&&&\\
D{\star D} P^\mu - {\star C^\mu\!_\sigma}\wedge P^\sigma &;&
D{\star C^\mu\!_\nu}
- {\star D} P^\mu\wedge \theta_\nu+ &;& -{\star C^\mu\!_\sigma}\wedge
\theta^\sigma\\
&&+\theta^\mu\wedge \star D P_\nu&&\\&&&&\\
0 &;&D{\star D} P_\nu +  P_\sigma\wedge \star C^\sigma\!_\nu &;&
-{\star D} P_\sigma\wedge \theta^\sigma
 \end{array}\right)\nonumber\\&&\label{J1}
\end{eqnarray}
The Greek indeces  are lowered and raised by
$\eta_{\mu\nu}$ and $\eta^{\mu\nu}$, respectively. In particular
\be \theta_\nu\ =\ \eta_{\nu\alpha}\theta^\alpha \ee
above. The differential three--form components of the  matrix
(\ref{J1}) will be decomposed in the basis
\be \star \theta_\mu\ :=\
\frac{1}{3!}\epsilon_{\mu\alpha\beta\gamma}\theta^\alpha\wedge \theta^\beta
\wedge \theta^\gamma, \ee
which satisfies the following identity, true for every
differential 3-form $W=W^\mu \star \theta_\mu$,
\begin{eqnarray*}
 \theta^\nu\wedge W  &=&\ W^\nu\vol,\\
\vol\ &=& \frac{1}{4!}\epsilon_{\alpha\beta\gamma\delta}\theta^\alpha\wedge
\theta^\beta\wedge \theta^\gamma\wedge \theta^\delta\ :=\ \theta^1\wedge \theta^2\wedge
\theta^3\wedge \theta^4.\ \end{eqnarray*}

Begin with the computation of the components ${J_{(C)}}^\mu_0$,
$\mu=1,...,4$ of the Yang--Mills current $J_{(C)}$ (that is the middle 4
terms of the zeroth column of $J_{(C)}$). They are given by

\begin{eqnarray}
&& \theta^\alpha\wedge D{\star D} P_\nu = \theta^\alpha\wedge
D{\star(D_\gamma P_{\nu\beta}\,\theta^\gamma\wedge \theta^\beta)} =
\frac{1}{2}\,\theta^\alpha\wedge D(D_\rho P_{\nu\sigma}\,\nonumber\\
&&\eta^{\rho\sigma}\!_{\gamma\beta}\,\theta^\gamma\wedge \theta^\beta)
=\frac{1}{2}\theta^\alpha\wedge (D_\delta D_\rho P_{\nu\sigma}\,
\epsilon^{\rho\sigma}\!_{\gamma\beta}\,\theta^\delta\wedge
\theta^\gamma\wedge
\theta^\beta)=\nonumber\\
&&=-\frac{1}{2} D_\delta D_\rho P_{\nu\sigma}\,
\eta^{\rho\sigma}\!_{\gamma\beta}\,\eta^{\alpha\delta\gamma\beta}\,
\textrm{vol} =  D_\delta D_\rho P_{\nu\sigma} (\eta^{\rho\alpha}\,
\eta^{\sigma\delta} - \nonumber\\
&&-\eta^{\rho\delta}\,\eta^{\sigma\alpha})\,\textrm{vol} = (
D^\sigma D^\alpha P_{\nu\sigma} -  D^\sigma D_\sigma
P_\nu\!^\alpha)\,\textrm{vol}
\end{eqnarray}

By a similar calculation we get
\begin{eqnarray}
&&\theta^\alpha\wedge P_\sigma\wedge{\star C^\sigma\!_\nu}
=-(P^{\sigma\beta}\,C_{\sigma\nu\beta}\!^\alpha)\,\textrm{vol}.
\end{eqnarray}
Therefore,
 \be {J_{(C)}}^\mu_0\ =\ D{\star D} P^\mu +  P_\sigma\wedge
\star C^{\sigma\nu} \ =\ B^{\mu\alpha}\,\star \theta_\alpha \ee
where $B^{\mu\alpha}$ is the Bach tensor that is, \be
B^{\mu\alpha}\ =\ 2 D_\sigma D^{[\alpha}\,P^{\sigma]\mu} -
2P^{\beta\sigma} \,{{{C_{\sigma}}^\mu}_\beta}^\alpha. \ee By the
symmetry of $J_{(C)}$, \be {J_{(C)}}^0_\nu\ =\ {B_\nu}^\alpha\star \theta_\alpha. \ee
Surprisingly, it turns out that all the other elements of $J_{(C)}$
identically vanish. Indeed, we find out, that the left-upper
corner component of $J_{(C)}$ is
\be {J_{(C)}}^0_0\ =\ \frac{1}{2}D_\nu G^{\nu\alpha}\,\star \theta_\alpha \ee
where
\be G^{\nu\alpha}\ =\ R^{\nu\alpha}\ -\
\frac{1}{2}R\eta^{\nu\alpha} \ee
is the Einstein tensor, hence $D_\nu G^{\nu\alpha}=0$.

To evaluate the middle block ${J_{(C)}}^\mu_\nu$, $\mu,\nu=1,...,4$, of
the Yang--Mills current $J_{(C)}$  we apply the tensor identity given by
the corresponding middle block of the conformal Bianchi identity
\be D\Omega_{(C)}\ =\ d\Omega_{(C)} + \omega_{(C)}\wedge\Omega_{(C)}-
\Omega_{(C)}\wedge \omega_{(C)}\ =\ 0, \ee and the known symmetry of the
Weyl tensor \be \star {C^\mu}_\nu\ =\
\frac{1}{2}{{{\eta^{\mu}}_\nu}_\gamma}^\delta\,{C^\gamma}_\delta.
\ee
The result is
\bd {J_{(C)}}^\mu_\nu\ =\ \left(-\frac{1}{2}\eta^{\alpha\mu}D_\delta
{G^\delta}_\nu + \frac{1}{2}\delta^\alpha_\nu D_\delta
G^{\delta\mu}\right)\,\star \theta_\alpha\ =\ 0. \ed

In conclusion, we proved that the Yang--Mills current of the Cartan
normal conformal connection  $\omega_{(C)}$ in the natural gauge
corresponding to a co-frame $(\theta^1,...,\theta^n)$  has the following
form,
\begin{equation}\label{theresult}
 J_{(C)} = \left(\begin{array}{ccc}
0 & 0 & 0 \\
B^{\mu\alpha}\star \theta_\alpha & 0 & 0 \\
0 & {B_\nu}^\alpha \star \theta_\alpha & 0 \end{array}\right)
\label{eq:jmatrix}
\end{equation}
where $B_{\mu\alpha}\theta^\mu\otimes \theta^\alpha$ is the Bach tensor of
the metric tensor $g=\eta_{\mu\nu}\theta^\mu\otimes \theta^\nu$.

 \textbf{Remark } A straightforward
consequence of (\ref{eq:jmatrix}) is the equivalence of the Bach
equation for conformal metric
\begin{displaymath}
B_{\mu\nu}=0
\end{displaymath}
and the Yang--Mills equation imposed on the Cartan normal
conformal connection \cite{merkulov}.
\begin{displaymath}
D\star\Omega_{(C)} = 0
\end{displaymath}

\subsection{In terms of the geometric definition}
We now turn to the geometric approach to the CNC
connection of Sec. \ref{2.2}. The Hodge star can be naturally
lifted to $\hstar$ acting on every horizontal\footnote{That is,
differential 2-forms such that contracted with every vector
tangent to a fiber of the bundle give zero.} 2-form $\hOmega'$
defined on $P(C)$. At $p\in P(C)$
\be \hstar\hOmega'\ :=\ \pi^*\star\sigma^*\hOmega',
 \ee
where $\sigma$ is any section of $P(C)$ containing $p$ in its
image. Notice, that  the curvature $\hOmega$ of a Cartan
connection is a horizontal 2-form. We define the {\it Yang--Mills
current} $\hJ_{(C)}$ of $\homega$ in the following way,
\be \hJ_{(C)}\ :=\ \hD\hstar\hOmega_{(C)}\ =\ d\hstar\hOmega_{(C)} +
\homega_{(C)}\wedge \hstar\hOmega_{(C)} - \hstar\hOmega_{(C)}\wedge
\homega_{(C)}.
\label{eq:current_g}\ee
The current $J_{(C)}$ introduced in the previous subsection  is related
to $\hJ_{(C)}$ by the natural sections of $P(C)$ defined in Sec.
\ref{2.2}. Let $\sigma_N$ be a natural section of $P(C)$
corresponding to the co-frame $(\theta^1,...,\theta^n)$ used in the
definition of $J_{(C)}$ above. Then
\be J_{(C)}\ =\ \sigma_N^*\hJ_{(C)}.\  \label{eq:current_g2}\ee
Therefore our result of the previous subsection seems to refer to
the natural gauge only. However, for a general (local) section
$\sigma:{\cal U}\rightarrow P(C)$ of the bundle $P(C)$ there is a
natural section $\sigma_N:{\cal U}\rightarrow P(C)$ such that
\be \sigma^*\he^\mu\  =\ \sigma_N^*\he^\mu\ =:\ \theta^\mu. \ee
Then, the corresponding pullbacks of $J_{(C)}$ are related by a gauge
transformation
\be \sigma^*J_{(C)}\ =\ h^{-1}\,\sigma_N^*J_{(C)}\, h \ee
where $h:{\cal U}\rightarrow H$ takes values in the subgroup of
the structure group $H$ given by  the first family of matrices in
(\ref{H}). The point is, that the formula (\ref{eq:jmatrix}) is
{\it invariant} with respect to those gauge transformations (see Appendix). This
observation allows us to strengthen the result: for every (local)
section $\sigma$ of $P(C)$, the pullback of the Yang--Mills
current of $\homega$ is
\begin{equation}
 \sigma^*J_{(C)} = \left(\begin{array}{ccc}
0 & 0 & 0 \\
B^{\mu\alpha}\star \theta_\alpha & 0 & 0 \\
0 & {B_\nu}^\alpha \star \theta_\alpha & 0 \end{array}\right).
\label{eq:jmatrix2}
\end{equation}
Since  $\hJ_{(C)}$ is a horizontal 3-form defined on $P(C)$, it follows
that $\hJ_{(C)}$ is
\begin{equation}
\hJ_{(C)} = \left(\begin{array}{ccc}
0 & 0 & 0 \\
B^{\mu\alpha}\widehat{\star \theta}_\alpha & 0 & 0 \\
0 & {B_\nu}^\alpha \widehat{\star \theta}_\alpha & 0
\end{array}\right), \label{eq:jmatrix3}
\end{equation}
where
\be \widehat{\star \theta}_\mu\  :=\
\frac{1}{3!}\eta_{\mu\alpha\beta\gamma}\he^\alpha\wedge \he^\beta
\wedge \he^\gamma, \ee
and $B_{\mu\nu}$ at $p\in P(C)$ is such that for every section
$\sigma$ such that $\sigma(m)= \pi(p)$, the tensor
$B_{\mu\nu}\sigma^*(\he^\mu\otimes\he^\nu)$ is the Bach tensor at
$m\in M$ of the metric tensor
$g=\eta_{\mu\nu}\sigma^*(\he^\mu\otimes\he^\nu)$.

\section{Reducible  non--conformally Einstein geometries}
Suppose that $[g]$ is a conformal geometry of the signature
$(p,q)$ and $\g\subset so(p+1,q+1)$ is a subalgebra. We say that
the CNC connection of is {\it reducible to} $\g$ in a neighborhood
of a point $m\in M$ if there is a choice of gauge (\ref{gauge})
such that the CNC connection 1-form $\omega_{(C)}$ takes values in
$T^*_mM\otimes\, \g$ at every point of some neighborhood of the
point  $m$. If the CNC connection of a given conformal geometry
$[g]$ is reducible to a single subalgebra $\g\subset so(p+1,q+1)$
at a neighborhood of every point $m\in M$, then we say that the
connection is reducible to $\g$, or just that the conformal
geometry $[g]$ is reducible to $\g$. In this case also the
Yang--Mills current takes values in the subalgebra. Due to
the result of the previous section, one may expect that the Bach
tensor of a reducible conformal geometry may be degenerate in some
way, depending on the Lie subalgebra $\g$. An example is the
family of the conformally Einstein geometries. Obviously, the Bach
tensor is zero at every point $m\in M$ in this case.
 Via the correspondence
between the twistorial \cite{rp} and the normal conformal Cartan
\cite{f} connections the reducibility of the first one implies the
reducibility of the second one. Therefore, another reducible
example is a conformal geometry admitting a covariantly constant
local twistor. All the conformal metric tensors of this property
and of the Lorenzian signature have been found in \cite{lewand1}.
In the generic (twisting local twistor field) case they were shown
to be the Fefferman geometries known in the theory of the
Cauchy--Riemann (CR) structures (3--dimensional in our case). 

The Yang--Mills
current corresponding to such CR structures has been derived in \cite{lewnur}. Below we use this
result to characterize the degeneracy of the Bach tensor of the
Fefferman metric. Indeed, it has only one independent real
component. On the other hand, the only Fefferman geometry which is
conformally Einstein is the flat geometry \cite{lewand2}. Putting
together those properties makes the class of the Fefferman metric
tensors a probable source of possible Bach flat metric tensors
which are {\it not} conformally Einstein.  Therefore, we consider
in this section the (source free) Yang--Mills equations imposed on
the CNC connection of the Fefferman metric
tensor. We derive all the homogeneous solutions.

\subsection{The Fefferman conformal geometries}
Note that given the CNC connection of an
initially {\it unknown} conformal geometry it is straightforward
to recover the geometry itself: four entries of the connection matrix 
constitute a null tetrad.

 On a 4-manifold $M$, we introduce
now a certain 1-form $\omega_{(CC)}$ taking values in $su(1,2)$,
next embed $\r :su(1,2) \rightarrow so(4,2)$ (the embedding $\r$ is defined
below) and notice that the
resulting 1-form $\r(\omega_{(CC)})$ {\it is} the CNC connection
1-form of the metric tensor we can read out from
$\r(\omega_{(CC)})$. The space-time $M$ and the 1-form $\omega_{(CC)}$
are constructed from a 3--Cauchy--Riemann (CR) geometry $(N,
[(\theta^1,\theta^3)])$; that is, from a real 3-manifold $N$ and an
equivalence class of pairs of one--forms $[(\mu,\lambda)]$ such
that
\begin{enumerate}
\item $\lambda$ is real, $\mu$ is complex
\item $\mu\wedge \overline{\mu}\wedge \lambda \neq 0$ everywhere
\item Two pairs $(\mu,\lambda)$ and $({\mu'},{\lambda'})$ are by
definition equivalent whenever
\be\label{eq:crgauge} {\lambda'} \ =\ f\, \lambda\ \qquad  {\mu'} = g\,
\mu \ +\ h\,\lambda \ee
where $f$ is an arbitrary real function while $g$ and $h$ are
arbitrary complex-valued functions.
\end{enumerate}

We assume throughout this work the following non-degeneracy
condition,
\begin{displaymath}\label{CRnondeg} \lambda\wedge d\lambda  \neq 0
\end{displaymath}
at every point of $N$, and given a CR geometry
$(N,[(\mu,\lambda)])$ we choose a representing pair
$(\mu,\lambda)$ such that the following normalization  condition
holds,
\be \label{CRnorm}\lambda\wedge d\lambda\ =\ i\mu\wedge
\bar{\mu}\wedge \lambda.\ee

We follow now the definition of the Fefferman metric formulated in \cite{bds}. 
Given a CR geometry $(N,[(\mu,\lambda)])$, and a representative
$(\mu,\lambda)$ the corresponding  4-space-time manifold $M$ is
defined as follows
\be M\ :=\  \{(x,e^{ir}\mu)\,|\,x\in N,\,\mu=\mu_{|x}, \,
r\in [0,2\pi] \}\label{eq:eir}\ee
The projection $\pi:M\rightarrow N$ is used to pull the forms
$\mu$ and $\lambda$ back into each point $(x,e^{ir}\mu)\in M$,
\be \omega_0\ :=\ \pi^*\lambda,\ \ {\omega_1}\ :=\
e^{ir}\pi^*\mu.\ee
The $su(1,2)$--valued connection 1-form $\omega_{(CC)}$ we introduce
on $M$ is the Cartan--Chern (CC) connection defined by the
following set of conditions:\footnote{This is again a `working'
definition, compatible with the working definition of the Cartan
normal conformal connection. For the full definition see
\cite{cartan,chernmosser}.}
\begin{equation}
\bs\omega_{(CC)} = \left(\begin{array}{ccc} \psi+i\theta^4&
\omega_1&-\omega_0 \\&&\\
i\,\overline{\omega_3} & -2i\theta^4 & i\,\overline{\omega_1}\\
&&\\
\omega_4 & -\omega_3 & -\psi+i\theta^4
\end{array}
\right)\label{eq:ccdef0}
\end{equation}
{\it where the real 1-forms $\psi,\, \theta^4,\, \omega_4$ and the
complex valued 1-form $\omega_3$ are such that the  CC curvature
takes the following form:}
\begin{equation}
\bs\Omega_{(CC)}=d\bs\omega_{(CC)} +
\bs\omega_{(CC)}\wedge\bs\omega_{(CC)}\ = \left(
\begin{array}{ccc}
0 & 0 & 0  \\&&\\
i\bar{R}\,\omega_1\wedge \omega_0 & 0 & 0 \\&&\\
\omega_0\wedge ({S}\,\omega_1 + \bar{S}\,\omega_2)  &
-R\,\bar{\omega_1}\wedge \omega_0  & 0
\end{array}\right) \label{eq:ccdef}
\end{equation}
{\it where  $R$ and $ S$ are any complex valued functions}.

In fact,  given the pair $(\mu,\lambda)$, the connection
$\omega_{(CC)}$ always exists,  and it is determined up to the gauge
transformation $\psi\mapsto \psi+ t\omega_0$, where $t\in \Real$
is arbitrary. An outline of the derivation and the result is given
in the Appendix. Clearly, the CC connection 1-form takes values in
the Lie algebra of matrices isomorphic with $su(1,2)$.
If we choose another pair $(\mu',\lambda')\in [(\mu,\lambda)]$,
the connection form transforms according to a law $\omega'_{(CC)}
= a^{-1}\omega_{(CC)}a + a^{-1}da$, where $a:M\rightarrow H$ is a 
suitable

The embedding $\r$ is the natural embedding of the algebra of the
$n$ by $n$ complex valued matrices into the algebra of the the
$2n$ by $2n$ real matrices: $\r(A)$ is given by replacing each
entry ${A^I}_J$ by a 2 by 2 block
\begin{displaymath}
\left(\begin{array}{cc}
\Re {A^I}_J & \Im {A^I}_J  \\
-\Im {A^I}_J & \Re {A^I}_J \\
\end{array}\right).
\end{displaymath}
Indeed, the resulting  1-form   $\r(\omega_{(CC)})$  takes the 
values in the Lie algebra $so(Q)$ defined in Section
2, where in this case
\be \eta\ =\ \left(\begin{array}{cccc}
0 & 0 & 0 & 1 \\
0 & 1 & 0 & 0  \\
0 & 0 & 1 & 0\\
1 & 0 & 0 & 0 \\
\end{array}\right).
 \ee
\be
\r(\omega_{(CC)}) = \left(\begin{array}{cccccc}
\psi&\theta^4 & \RE \omega_1 & \IM \omega_1 & -\omega_0 & 0 \\
-\theta^4 & \psi & -\IM \omega_1 & \RE \omega_1 & 0 & -\omega_0 \\
\RE \omega_3 & \IM \omega_3 & 0 & -2\theta^4 & \IM \omega_1 & \RE \omega_1 \\
-\IM \omega_3 & \RE \omega_3 & 2\theta^4 & 0 & -\RE \omega_1 & \IM \omega_1\\
\omega_4 &0&-\IM \omega_3 &-\RE \omega_3&-\psi&\theta^4 \\
0 & \omega_4 &\RE \omega_3 & -\IM \omega_3 & -\theta^4 &-\psi\end{array}\right)
\ee
Note that {\it if} $\r(\omega_{(CC)})$ {\it is} the CNC connection of some
metric tensor, then the corresponding co-frame
$(\theta^1,\theta^2,
\theta^3,\theta^4)$ and
the metric tensor $g=g_F$ can be immediately identified as
\be \theta^1\ :=-\omega_0,\quad \theta^2\ :=\RE\omega_1,\quad 
\theta^3\ :=\ \IM\omega_1\qquad
g_F\ :=\ \eta_{\mu\nu}\theta^\mu\otimes \theta^\nu
 \ee
where $\theta^4$ and $\eta$ coincide with the ones already defined in this
section. 
As before the dual basis of vectors at each point will be denoted by $(X_1,\ldots,X_4)$. The linear independence of the co-frame follows from
(\ref{eq:ccdef}) and (\ref{CRnondeg}).  
The resulting metric $g_F$
is {\it the Fefferman metric tensor} \cite{feff,bds}.

The signature of the metric is $(-+++)$. The metric $g_F$ is
determined by the pair of 1-forms $(\mu,\lambda)$ whereas it is
independent of the remaining ambiguity in the CC connection 1-form
$\omega$. 
The null vector field $X_4 = k$ is a Killing vector field for $g_F$.
It generates the flow $R_t \colon (x,e^{ir}\mu) \to (x,e^{i(r+t)}\mu)$. 
It can be checked by inspection that the definition of
the CC connection 1-form $\omega_{(CC)}$ implies that
$\r(\omega_{(CC)})$ {\it is} the CNC connection of the Fefferman
metric $g_F$,
\be \r(\omega_{(CC)})\ =\ \omega_{(C)}\ee

 The Fefferman metric $g'_F$ assigned to another pair
$(\mu',\lambda'=e^h\lambda)\in [(\mu,\lambda)]$ (remember about
the normalization (\ref{CRnorm})) and defined on $M'$ is related to
$g_F$ by the map $\varphi: (x,e^{ir}\mu) \mapsto
(x,e^{ir}\mu')$ \cite{feff,bds}
\be  \varphi^*\left( g'_F\right)\ =\ e^hg_F. \ee

Another useful technical
remark is that if we represent a given CR geometry by a pair
$(\mu,\lambda)$ such that
\be d\lambda\ =\ i\mu\wedge\bar{\mu}, \ee
and use $\psi=0$  (this combination of conditions is always
possible) then, the resulting CNC connection 1-form
$\r(\omega_{(CC)})$ comes out in the natural gauge (\ref{omegaN}).

The CNC curvature form is 
\begin{displaymath}
 \Omega_{(C)}  = 
 \left(\begin{array}{cccccc}0 & 0 & 0 & 0 & 0 &0 \\
0&0&0&0&0&0\\
\RE(R\omega_2) & \IM(R\omega_2) & 0&0&0&0\\
-\IM(R\omega_2)&\RE(R\omega_2) &0&0&0&0\\
-\RE(S\omega_1)&0&\RE(R\omega_2)&\IM(R\omega_2)&0&0\\
0&-\RE(S\omega_1)&-\IM(R\omega_2)\omega_0&\Re(R\omega_2)&0&0
\end{array}\right)\wedge\omega_0
\end{displaymath}
Note that the whole middle block vanishes iff $R$ vanishes, that is,
Fefferman metric is conformally flat iff $R=0$.

\subsection{The Bach tensor of the Fefferman metrics}
Combining the results of the previous and the current sections,
one can see that the Bach tensor of the the Fefferman metric is
given by the Yang--Mills current $J_{(CC)}$ of the CC connection
coupled with the corresponding Fefferman metric,
\begin{eqnarray*}
J_{(CC)}\ &=&\ D\star\Omega_{(CC)}\ =\ d\star\Omega_{(CC)} +
\omega_{(CC)}\wedge\star\Omega_{(CC)} - \star\Omega_{(CC)}\wedge \omega_{(CC)}
\\
J_{(CC)} &=& \left(\begin{array}{ccc}
0&0&0\\0&0&0\\\kappa&0&0\end{array}\right) \\
 J_{(C)}\ &=&\ \r(J_{(CC)}) =  \left(\begin{array}{cccccc} 0&0&0&0&0&0 \\
0&0&0&0&0&0\\0&0&0&0&0&0\\0&0&0&0&0&0\\\kappa&0&0&0&0&0\\
0&\kappa&0&0&0&0\end{array}\right)
\end{eqnarray*}

where 
\ba \kappa &=& 2\left(\omega_0\wedge d\IM(S\omega_1)\right)-2
\RE\left(R\bar\omega_1\wedge\bar\omega_3\right)\wedge\omega_0 =
\nonumber\\
&=& -2(\theta^1\wedge d\IM(S\theta^2))-
2\RE(R\theta^3\wedge\bar\omega_3)\wedge\theta^1
\label{eq:kappa1}\ea

As we know from Section 1, the Bach tensor may be calculated from 
the Yang--Mills current via 
\be
B^{\mu\alpha}\,\textrm{vol} = \theta^{\alpha}\wedge J_{(C)}\,^\mu_0
\ee
The only non--vanishing entry of the current is $J_{(C)}\,^4_0$, therefore
\be
B^{1 \alpha}=B^{2\alpha}=B^{3\alpha}=0 \qquad B^{44} \neq 0 \textrm{ in general}
\ee
Since $B^{\mu\nu}$ is a symmetric tensor, this result means that the
only non--vanishing component of the Bach tensor is $B^{44}$. 
In other words 
\be
B^{\mu\nu}\,X_{\mu}\otimes X_{\nu} = 
T \cdot X_4\otimes X_4 = 0
\ee
where $X_4$ is the null Killing field of the Fefferman 
metric
 and $T$ is a function which can be explicitly calculated using
(\ref{eq:kappa1}). Therefore, as it was anticipated in Introduction, the 
Bach tensor is degenerate.

{\bf Remarks. }
\begin{enumerate}
\item $J_{(C)}$ does not contain $\theta^4$ 
and therefore the contraction $i(X_4) J_{(C)}$ 
vanishes.

\item The Yang--Mills equations imposed on the normal Cartan connection of a
Fefferman metric proved to be equivalent to one single real equation
\be
\kappa = 0 \label{eq:kappa}
\ee
This fact was first pointed out in \cite{lewnur} but there was a misprint
in the explicit formula for (what we currently call) $\kappa$.
This is the reason why we state the correct formula here in detail.  
\end{enumerate}
 Since $X_4$ is a Killing vector field of the metric tensor,
the Yang--Mills equations for the CC connection amount to
a differential equation on the forms $\mu$ and $\lambda$ defined on $N$.  
In order to study it, in the next subsection 
we take the section   $\xi \colon N \ni {\cal O} \mapsto M$
defined  by $r=0$ in (\ref{eq:eir}), and  consider the pullbacks of the CC 
connection and curvature forms to the CR manifold $N$.
These pullbacks will be referred to as the  
\emph{CC connection and curvature forms on $N$} respectively.
 
\subsection{Homogeneous Fefferman geometries}

We now turn to a specific class of highly symmetric CR geometries.
We will assume the CR geometry $N$ admits a 3 dimensional group $G_3$ 
of symmetries. Such geometries are called \emph{homogeneous}. In this class 
we will solve the Yang--Mills equations for the CC connection.

We will use the following notation for $\lambda$ and $\mu$ defined on $N$:  
\ba
e^1 = -\mu = -\xi^*(\theta^2+i\theta^3)&& e^2 = -\bar\mu =
-\xi^*(\theta^2 - i\theta^3)\\
e^3 = \lambda = -\xi^*\theta^1&&  
\ea
and for the pullback of CC connection matrix elements
\ba
\omega_{III} = -\xi^*\omega_3 && \omega_{IV} = \xi^*\omega_4 \\
\omega_{II} = \xi^*(-\psi-\frac{3i}{2}\theta^4)
\ea
This choice is just a matter of convenience and was made to match
the notation of \cite{lewphd}.

The CC connection form on $N$ satisfies the same condition 
(\ref{eq:ccdef}) with CC curvature form on $N$ on the right hand side.
In the Appendix we present an outline of derivation of  
CC connection one--form on $N$ of a given CR geometry.

The forms $e^1$, $e^2$ and
$e^3$ on a homogeneous CR manifold may be chosen in such a way that their exterior derivatives were decomposable
with constant coefficients in the
basis of wedge products of $e^i$ \cite{nurtaf}.
The forms $e^1$, $e^2$ and $e^3$ are then the left--invariant one--forms on the
homogeneous space of $G_3$.
\begin{eqnarray*}
de^3 &=& A\,e^3\wedge e^1 + \bar A\,e^3\wedge e^2 + iB\,e^1\wedge e^2 \\
de^1 &=& C\,e^3\wedge e^1 + D\,e^3\wedge e^2 + E\,e^1\wedge e^2\\ \\
A,C,D,E \in \mathbf{C} &\qquad& B \in \mathbf{R}
\end{eqnarray*}

This condition doesn't fix the forms completely: a residual gauge freedom 
remains like
in (\ref{eq:crgauge}) with $f$, $g$ and $h$ \emph{constant numbers}.
We may reduce this freedom by a further requirement that the forms satisfy
\begin{eqnarray}
de^3 \wedge e^3 &=& i\,e^1\wedge e^2\wedge e^3 \label{eq:cond3}\\
de^1 \wedge e^1 &=& 0 \label{eq:cond1}
\end{eqnarray}
or
\begin{eqnarray}
de^3 &=& A\,e^3\wedge e^1 + \bar A\,e^3\wedge e^2 + i\,e^1\wedge e^2 \nonumber\\
de^1 &=& C\,e^3\wedge e^1 + E\,e^1\wedge e^2 \label{eq:roz}
\end{eqnarray}
This can always be done provided that the CR structure is non--degenerate. 
Indeed, if 
\ba
de^3\wedge e^3 &=& iP\, e^1\wedge e^2\wedge e^3\\
de^1\wedge e^1 &=& Q\, e^1\wedge e^2\wedge e^3
\ea
then the forms
\ba
 e_{New}^1 &=& e^1 + h\,e^3\\
 e_{New}^3 &=& \frac{1}{P}\,e^3
\ea
satisfy the gauge conditions iff $h$ is such that
\be
0=de^1\wedge e^1 + h(de^1\wedge e^3 + de^3\wedge e^1) + h^2 de^3\wedge e^3
\label{eq:quadratic}
\ee
(\ref{eq:quadratic}) is a complex quadratic equation and therefore always
has a solution.  
Moreover in the generic case it has two solutions. This gives rise to an
additional discrete gauge
transformation of changing of the root of (\ref{eq:quadratic}) and causes
some complications.

Calculating the CC connection and its curvature is quite easy due to the fact that $A$, $C$ and $E$ are constant.
By plugging (\ref{eq:roz}) into (\ref{eq:pierwszy})--(\ref{eq:ostatni}) (Appendix)
we get the connection
 form elements
\begin{eqnarray}
\omega_{II\,2} &=& -E \\
\bar\omega_{II\,1} &=& -\bar E \\
\omega_{III\,2} & =& 0 \\
\bar\omega_{III\,1} & =& 0 \\
\omega_{II\,1} &=& -A + \bar E \\
\bar\omega_{II\,2} &=& -\bar A + E \\
\omega_{II\,3} &=& \frac{i\beta}{4}
\end{eqnarray}
where we introduced for convenience
\begin{displaymath}
\beta =
3\,\textrm{Im}\,C - \textrm{Re}\,EA + 2|E|^2
\end{displaymath}
\begin{eqnarray}
\omega_{III\,1}&=&-C+\frac{i\beta}{4} \\
\bar\omega_{III\,2}&=&-\bar C-\frac{i\beta}{4} \\
\omega_{III\,3}&=&\bar A\left(\frac{i\bar C}{3} - \frac{2iC}{3} -
\frac{\beta}{6}\right) \\
\omega_{IV\,1} &=&\frac{C}{3}(A+\bar E)+\frac{i\beta A}{12} \\
\omega_{IV\,2} &=&\frac{\bar C}{3}(\bar A+E)-\frac{i\beta\bar A}{12} \\
 \omega_{IV\,3}&=&\frac{i\beta C}{2} - C^2 + \frac{\beta^2}{16} + (\bar E+A)\,
\left(\frac{i\bar C\bar A}{3}-
\frac{2i\bar C E}{3}-\frac{\beta\bar A}{6}\right)
\end{eqnarray}
Analogous expressions for the curvature
\begin{eqnarray}
 R &=& \bar A\left(\frac{i\bar C}{3} - \frac{2iC}{3} - \frac{\beta}{6}\right)
\left(-2\bar A + E\right) \label{eq:R2}\\
 S&=&C\left(\frac{C}{3}(A+\bar E) + \frac{iA\beta}{12}\right) +
2A\left(\frac{i\beta C}{6} - C^2 + \frac{\beta^2}{16}+ \right.\nonumber\\
&+&
\left.(\bar E+A)\left(\frac{i\bar C\bar A}{2} - \frac{2i\bar C E}{3} -
\frac{\beta\bar A}{6}\right)\right) + \nonumber\\
&+&iA\left(-C+\frac{i\beta}{4}\right)
\cdot\left(-\frac{iC}{3}+\frac{2i\bar C}{3}-\frac{\beta}{6}\right)
\label{eq:S2}
\end{eqnarray}
By applying this we find out that (\ref{eq:kappa}) is equivalent to
\begin{equation}
0= S\,\left(\frac{1}{2}E + \bar A\right)
\end{equation}
We conclude that the CC connection satisfies YM iff $E = -2\bar A$ or $ S=0$.
Before we proceed we will investigate the second condition in details.

In the case of $ S=0$ the Bianchi identity
\begin{displaymath}
 d\Omega_{(CC)} + 
\omega_{(CC)}\wedge\Omega_{(CC)} - \Omega_{(CC)}\wedge\omega_{(CC)}=0
\end{displaymath}
implies that
\begin{displaymath}
\bar R\,(2E+\bar A) = 0
\end{displaymath}
Once again we get two (perhaps overlapping) cases: $ R = 0$ i
$E=-1/2\,\bar A$. The first one clearly corresponds vanishing $\Omega_{(CC)}$. 
As we concluded in the previous subsection this means that the Weyl tensor of the Fefferman metric is equal to 0 and hence the metric is conformally flat.

We may summarize the result as follows: (\ref{eq:kappa})
reduces to one algebraic equation involving $A$, $C$ and $E$.
Homogeneous CR geometries satisfying (\ref{eq:kappa}) fall into three categories:
\begin{enumerate}
\item $ S = 0$ i $ R = 0$ (the flat case)
\item $ S = 0$ i $E = -1/2\bar A$
\item $E = -2\bar A$
\end{enumerate}
Only the latter two may actually contain non--trivial (non--conformally flat)
solutions.

The final step involves using a complete classification of homogeneous, three--dimensional
CR geometries as presented in \cite{nurtaf}. The classification is based on Bianchi classification
of the group $G_3$. Before we apply it we must impose our gauge conditions to the forms presented
in \cite{nurtaf} and calculate $A$, $C$ and $E$. We will just state the results here:
\vskip 0.5cm
\begin{tabular}{|l|c|c|c|}
\hline
\emph{Type} & \emph{A} & \emph{C} & \emph{E} \\
\hline
II [A] & 0 & 0 & 0 \\
\hline
IV [F] & $-i/2$ & 0 & $i/2$\\
\hline
VI$_h$ including VI$_0$ and III [E,B] &
$ib/2$ & 0 & $i/2$ \\
\hline
VII$_h$ including VII$_0$ & $b+i$ & $2(b+i)$ & $b+i$ \\
\hline
IX [D,L] and VIII [C,K] & $ik/2$ & $i(k^2\pm 1)/2$ & $ik/2$ \\
\hline
\end{tabular}
\vskip 0.5cm
Some of types involve a one--parameter family of CR geometries. It's worth
mentioning that the CR structures of type VI corresponding to $b$ and
$b^{-1}$ are isomorphic, as well as $b$ and $-b$ in type VII and
$k$ and $-k$ in type IX. This is due to the previously mentioned discrete gauge symmetry.

Simple inspection of the table convinces us that
\begin{itemize}
\item     type II is the flat case
\item     $E = -2\bar A$ are satisfied only by types types II and
VI$_h$ with $b=\frac{1}{2}$
\item     $E = -1/2\bar A$ are satisfied only by types II and VI$_h$ with $b=2$
\end{itemize}
Therefore type VI$_h$ with $b=\frac{1}{2}$ and type VI$_h$ with $b=2$
are the only solutions of the YM equation
which may prove non--flat. We now check if this is really the case.
the general formula for the curvature coefficients of CC connection
associated with type VI$_h$ homogeneous CR geometry:
\begin{displaymath}
 R = -\frac{ib}{2}\left(-\frac{b+2}{24}\right)\left(ib+\frac{ib}{2}
\right)=-\frac{b(b+2)(b+1/2)}{48}
\end{displaymath}
\begin{eqnarray*}
S&=&2A\left(\frac{\beta^2}{16}+(\bar E+A)\left(-\frac{\beta \bar A}{6}
\right)\right)+iA\cdot\frac{i\beta}{4}\left(-\frac{\beta}{6}\right)=\\
&=&\frac{\beta A}{3}\left(\frac{\beta}{2}-\bar E\bar A-|A|^2\right)=
-\frac{b(b+2)}{24}\left(-\frac{b^2}{4}+\frac{3}{8}b+ \frac{1}{4}
\right)=\\
&=&\frac{b(b+2)(b-2)(b+1/2)}{96}
\end{eqnarray*}
$ S$ doesn't vanish for either $b=\frac{1}{2}$ or $b=2$.
Both coefficients of the curvature vanish for $b=-\frac{1}{2}, -2$ and 0.
However, as we have shown in the appendix, the CR structures $b$ and $b^{-1}$
are in fact isomorphic. This is due to the discrete gauge symmetry mentioned
in section the Appendix.

Our consideration can now be summarized:
in the type VI$_h$ family of CR structures the YM equation is satisfied for
\begin{itemize}
\item $b=0$ -- flat
\item $b=-2$ and $b=-\frac{1}{2}$ (isomorphic) -- flat
\item $b=2$ i $b=\frac{1}{2}$ (isomorphic) -- the only non--flat solutions
\end{itemize}

We did not exclude the existence of trivial solutions among other types
of homogeneous CR geometries.

Using the coordinates on $N$ introduced in \cite{nurtaf} for each type
of homogeneous CR geometry 
we can write the Fefferman metrics corresponding to non--trivial solutions
\begin{itemize}
\item[--]$b=\frac{1}{2}$:
\begin{equation}
g=dx^2+dy^2+\frac{4}{3}\left(y^{3/2}du-dx\right)
\left(y\,dr + \frac{5}{36}\,y^{3/2}\,du+\frac{25}{36}\,dx\right)
\label{eq:nnn}
\end{equation}
\item[ii]$b=2$:
\begin{equation}
g=dx^2+dy^2 + \frac{2}{3}\left(y^3\,du - dx\right)\left(y\,dr +
\frac{1}{9}\,y^3\,du+\frac{11}{9}\,dx\right) \label{eq:nurpleb}
\end{equation}
\end{itemize}
(\ref{eq:nnn}) and (\ref{eq:nurpleb}) are conformally equivalent as they are constructed from
isomorphic CR geometries. Indeed, it's straightforward to verify that the coordinate change
\begin{eqnarray*}
dx &=& -\frac{1}{2}\,d\tilde u \\
dy &=& -\frac{1}{2}\,y^{-3/2}\,d\tilde y \\
du &=& -\frac{1}{2}\,d\tilde x
\end{eqnarray*}
transforms (\ref{eq:nurpleb})
into (\ref{eq:nnn}) up to a conformal factor.

Metric (\ref{eq:nurpleb}) as a Bach and non-conformally Einsteinian metric
appeared for the first in \cite{nurpleb}, it was derived
however in a different way.

The results of this chapter were verified using the www version of
symbolic calculations program GRTensor.
\bigskip

\noindent{\bf Acknowledgments.} We thank Pawel Nurowski, Ted Newman, 
Roger Penrose and Andrzej Trautman for the discussions and  remarks.
MK would also like to thank Pawel Nurowski for introduction
in the field of Cartan connections. 

The work was partially supported  by the Polish KBN grant  
number 2 PO3B 12724.   

\appendix
\section{Notation and identities}
\subsection{Bach tensor and Bianchi identities}
Apart from the standard Weyl and Ricci tensors the following
Riemann geometry objects have been used
\begin{displaymath}
P_{\mu\nu} = \frac{1}{12}\,R\,g_{\mu\nu} - \frac{1}{2}\,R_{\mu\nu}
\end{displaymath}
and the \emph{Bach tensor}
\begin{equation}
B_{\mu\nu} = 2\nabla^\sigma\nabla_{[\nu}P_{\sigma]\mu} -
P_{\rho\sigma} C_\mu\!^\sigma\!_\nu\!^\rho \label{eq:bach1}
\end{equation}
\subsection{Volume and the antisymmetric symbol}
We define the volume form in an orthonormal frame by
\begin{displaymath}
\textrm{vol} = \theta^0\wedge\theta^1\wedge\theta^2\wedge\theta^3 =
\frac{1}{24}\,\epsilon_{\mu\nu\alpha\beta}
\theta^\mu\wedge\theta^\nu\wedge\theta^\alpha\wedge\theta^\beta
\end{displaymath}
where the antisymmetric symbol $\epsilon$ is in our convention
\begin{displaymath}
\epsilon_{0123} = 1 \qquad \epsilon^{0123} = -1
\end{displaymath}
With this setup we have the following identities
\begin{eqnarray}
\theta^\mu\wedge\theta^\nu\wedge\theta^\alpha\wedge\theta^\beta&=&
-\epsilon^{\mu\nu\alpha\beta}\,\textrm{vol} \label{eq:olda5}\\
\frac{1}{2}\,\epsilon^{\mu\nu\rho\sigma}\,
\epsilon^{\alpha\beta}\!_{\rho\sigma}
&=& - \eta^{\mu\alpha}\eta^{\nu\beta} + \eta^{\mu\beta}\eta^{\nu\alpha}
\label{eq:olda6}\\
\frac{1}{6}\,\epsilon_{\mu\nu\rho\sigma}\,\epsilon^{\alpha\nu\rho\sigma} &=&
-\eta^{\mu\alpha} \label{eq:olda7}
\end{eqnarray}

\subsection{The adjoint action of $SO(2,4)$ on ${so}(2,4)$}
Taking into account the definition of the form $Q$
every $A \in {so}(2,4)$ has the form of
\begin{displaymath}
A = \left(\begin{array}{ccc}
C & U_\nu & 0 \\ V^\mu & L^\mu\!_\nu & U^\mu \\
0 & V_\nu & -C \end{array}\right)
\end{displaymath}
The adjoint action of 3 subgroups of $SO(1,3)$ may easily be
computed. 
\begin{itemize}
\item \emph{dilations}
\begin{equation}
\begin{array}{rcl}
U^\mu & \mapsto & c\cdot U^\mu \\
V_\nu & \mapsto & c^{-1}\cdot V_\nu
\end{array} \label{eq:dyll}
\end{equation}
other elements do not change
\item \emph{Lorentz transformations}
\begin{equation}
\begin{array}{rcl}
U^\mu & \mapsto & \Lambda^\mu\!_\nu U^\nu \\
V_\nu & \mapsto & (\Lambda^{-1})\,^\mu\!_\nu\,V_\mu \\
L^\mu\!_\nu & \mapsto & \Lambda^\mu\!_\rho\,L^\rho\!_\sigma\,
(\Lambda^{-1})^\sigma\!_\nu
\end{array} \label{eq:lorr}
\end{equation}
other elements do not change
\item \emph{M\"obius transformations}
\begin{equation}
\begin{array}{rcl}
C & \mapsto & C - U^\alpha\,\xi_\alpha \\
L^\mu\!_\nu & \mapsto & L^\mu\!_\nu + \xi^\mu U_\nu - \xi_\nu U^\mu \\
V_\nu & \mapsto & V_\nu + (L^\mu\!_\nu + \delta^\mu\!_\nu\,
(C - \xi^\alpha U_\alpha))\xi_\mu
\end{array} \label{eq:ady}
\end{equation}
other elements do not change
\end{itemize}
The curvature form $\Omega_{(C)}$ is a type \textit{ad} form. The transformation law
of the Cartan normal connection form is more complicated, but
involves adjoint action on $so(2,4)$ too. Therefore
 the equations above may be used to derive the transformation laws for
geometric objects of conformal geometry: $P_{\mu\nu}$, $B_{\mu\nu}$ and 
$C^\mu\!_{\nu\rho\sigma}$.

\subsection{Admissible transformations of $e^1$ and $e^3$
for homogeneous CR geometries with the gauge conditions imposed}

With the conditions (\ref{eq:cond3}) and (\ref{eq:cond1}) imposed
there remains one--parameter gauge freedom in the choice of
forms $\{e^1,e^3\}$
\begin{eqnarray*}
e^1&\mapsto& g\cdot e^1 \\
e^3&\mapsto& |g|^2\,e^3
\end{eqnarray*}
which results in changing the parameters $A$, $C$ and $E$:
\begin{displaymath}
A\mapsto\frac{A}{g}\qquad C\mapsto\frac{C}{|g|^2}\qquad E\mapsto\frac{E}
{\bar g}
\end{displaymath}
We also have an unexpected transformation of changing the
root of the quadratic equation (\ref{eq:quadratic}):
\begin{eqnarray*}
e^1&\mapsto& e^1 + i(E - \bar A)\,e^3\\
e^3&\mapsto& e^3
\end{eqnarray*}
with the resulting change of structure constants
\begin{displaymath}
A\mapsto\bar E\qquad C\mapsto C-2\textrm{Im} EA\qquad E\mapsto\bar A
\end{displaymath}
The transformation is clearly an involution.
This fact lies behind the isomorphisms of homogeneous structures
of various types mentioned in the last section.

\subsection{Derivation of the elements of the CC connection form on $N$ from the exterior derivatives of $e^1$ and $e^3$}

$\omega_{II}$, $\omega_{III}$, $\omega_{IV}$ and the curvature can be expressed in the terms of $e^1$ and $e^2$.
The method of calculation is taken from \cite{lewphd}.

The condition (\ref{eq:ccdef}) yields six equations
\begin{eqnarray}
de^1 &=&  \omega_{II} \wedge e^1 + \omega_{III} \wedge e^3
\label{eq:cc1}\\
de^2 &=& \bar\omega_{II} \wedge e^2 + \bar\omega_{III} \wedge e^3
\label{eq:cc2}\\
de^3 &=& i\,e^1\wedge e^2 + (\omega_{II} + \bar\omega_{II})\wedge e^3
\label{eq:cc3}\\
d\omega_{II} &=& 2i\,e^1\wedge \bar\omega_{III} + i\,e^2\wedge \omega_{III} +
\omega_{IV} \wedge e^3 \label{eq:cc4}\\
d\omega_{III}&=&\omega_{IV}\wedge e^1 + \omega_{III}\wedge \bar\omega_{II} +
 R\,e^2\wedge e^3
\label{eq:cc5}\\
d\omega_{IV}&=&
\omega_{IV}\wedge(\omega_{II} + \bar\omega_{II}) -
i\bar\omega_{III}\wedge\omega_{III}-\nonumber\\
&&  S\,e^1\wedge e^3 -
\bar S\,e^2\wedge e^3 \label{eq:cc6}
\end{eqnarray}
Note that (\ref{eq:cc1}) is the complex conjugate of (\ref{eq:cc2}).

We will first consider the equations (\ref{eq:cc1})--(\ref{eq:cc3}). By taking their relevant components we get
\begin{eqnarray}
\omega_{II\,2} &=& -(de^1)_{12} \label{eq:pierwszy}\\
\bar\omega_{II\,1} &=& (de^2)_{12}\nonumber \\
\omega_{III\,2} &=& (de^1)_{23} \nonumber\\
\bar\omega_{III\,1} &=& (de^2)_{123} \nonumber
\end{eqnarray}
and two relations
\begin{eqnarray*}
(de^3)_{13} &=& \omega_{II\,1} + \bar\omega_{II\,1}\\
(de^3)_{23} &=& \omega_{II\,2} + \bar\omega_{II\,2}
\end{eqnarray*}
which we may combine with the previous results and find out that
\begin{eqnarray*}
\omega_{II\,1} &=& (de^3)_{13} - (de^2)_{12} \\
\bar\omega_{II\,2} &=& (de^3)_{23} + (de^1)_{12}
\end{eqnarray*}
Finally we note the following relations
\begin{eqnarray}
(de^1)_{13} &=& -\omega_{II\,3} + \omega_{III\,1} \label{eq:tozs1}\\
(de^2)_{23} &=& -\bar\omega_{II\,3} + \bar\omega_{III\,2} \label{eq:tozs2}
\end{eqnarray}
which we will use further on.

We calculate both sides of (\ref{eq:cc4}) keeping in mind that $\omega_{IV}$ and $e^3$ are real
\begin{displaymath}
2i\textrm{ Im }d\omega_{II} \equiv d\omega_{II} - d\bar\omega_{II} =  3i\left(e^1\wedge\bar\omega_{III} + e^2\wedge\omega_{III}\right)
\end{displaymath}
or
\begin{equation}
d\textrm{ Im }\omega_{II} = \frac{3}{2}(e^1\wedge\bar\omega_{III} + e^2\wedge\omega_{III}) \label{eq:4.18}
\end{equation}
The left hand side
\begin{displaymath}
\textrm{l.h.s.}=d\textrm{ Im }(\omega_{II\,1}e^1 + \omega_{II\,2}e^2) + \textrm{ Im }(d\omega_{II\,3}\wedge e^3 +
\omega_{II\,3} de^3)
\end{displaymath}
The first term inside the second bracket does not contain $e^1\wedge e^2$. Therefore
\begin{displaymath}
\textrm{(l.h.s.)}_{12} = d(\textrm{ Im }(\omega_{II\,1}e^1 + \omega_{II\,2}e^2)_{12} +
i\omega_{II\,3}
\end{displaymath}
\begin{displaymath}
(\textrm{r.h.s.})_{12} = \frac{3}{2}(\bar\omega_{III\,2}-\omega_{III\,1})
\end{displaymath}
The latter equation after substituting (\ref{eq:tozs1}) and (\ref{eq:tozs2}) takes the form of
\begin{displaymath}
(\textrm{r.h.s.})_{12} = \frac{2}{3}\left((de^2)_{23} - (de^1)_{13}\right) - 3i \textrm{ Im }\omega_{II\,3}
\end{displaymath}
Combining both sides yields an equation for the imaginary part of $\omega_{II\,3}$ expressed in the terms of quantities
we already know
\begin{displaymath}
i\textrm{ Im }\omega_{II\,3} = \frac{3}{8}\left((de^2)_{23}-(de^1)_{13}\right) - \frac{1}{4}\left(d\textrm{ Im }(\omega_{II\,1}
e^1 +\omega_{II\,2}e^2)\right)_{12}
\end{displaymath}
The real part of $\omega_{II\,3}$ can be assumed to be equal to naught. Hence we have got the whole form $\omega_{II}$
expressed in the terms of external derivatives of $e^1$ and $e^3$.
This allows us to calculate the whole $\omega_{III}$: components 1 and 2 from (\ref{eq:tozs1}) and (\ref{eq:tozs2}), while
the third component from (\ref{eq:4.18})
\begin{displaymath}
\omega_{III\,3} = \frac{2}{3}(d\textrm{ Im }\omega_{II})_{23}
\end{displaymath}
We now take the 13 component of (\ref{eq:cc4})
\begin{displaymath}
\omega_{IV\,1}=(d\omega_{II})_{13} - 2i \bar\omega_{III\,3}
\end{displaymath}
$\omega_{IV}$ is real, therefore
\begin{displaymath}
\omega_{IV\,2} = \overline{\omega_{IV\,1}}
\end{displaymath}
The component 13 of (\ref{eq:cc5}) yields
\begin{displaymath}
\omega_{IV\,3} = (\omega_{III}\wedge\bar\omega_{II})_{13} - (d\omega_{III})_{13}
\end{displaymath}
Finally we can use (\ref{eq:cc5}) and (\ref{eq:cc6}) to calculate the curvature coefficients
\begin{eqnarray}
 R &=& (d\omega_{III})_{23} - (\omega_{III}\wedge\bar\omega_{II})_{23}
\nonumber\\
 S &=& (-d\omega_{IV})_{13}+(\omega_{IV}\wedge(\omega_{II}+\bar\omega_{II}))_{13} \label{eq:ostatni}
\end{eqnarray}


\begin{thebibliography}{99}
\bibitem{rp} Rindler W and Penrose R 1986
\emph{Spinors and Spacetime v. 1 \& 2 } (Cambridge: Cambridge University Press)
\bibitem{kob} Kobayashi S 1972 \emph{Transformation Groups In Differential
Geometry} (W\"urzburg: Springer-Verlag)
\bibitem{nurtaf}Nurowski P and Tafel J 1988 \emph{Lett. Math. Phys.} \textbf{15} 31
\bibitem{lewand1}Lewandowski J 1991 \emph{Class. Quantum Grav.} \textbf{8} L11-L17
\bibitem{nurpleb}Nurowski P and Pleba\'nski J 2001 \emph{Class. Quantum Grav.} \textbf{18} 
341
\bibitem{bds} Burns Jr D, Diederich K and Shnider S 1977 \emph{Duke Math. J.} \textbf{44}
407
\bibitem{lewnur}Lewandowski J and Nurowski P  \emph{J. Geom. Phys.} \textbf{7} 63
\bibitem{lewphd}Lewandowski J 1989 \emph{PhD Thesis} (University of Warsaw)
\bibitem{lewand2}Lewandowski J 1998 \emph{Lettr. Math. Phys.} \textbf{15} 129
\bibitem{feff} Fefferman C L 1976 \emph{Ann. Math.} \textbf{103} 395
\bibitem{merkulov} Merkulov S A 1984 \emph{Class. Quantum Grav.} \textbf{1} 349
\bibitem{lionel} Mason L 1986 \emph{Twistor in Curved Spacetime} Oxford D. Phil.Thesis  
\bibitem{knt} Kozameh C N, Newman E T and Tod K P 1985
\emph{Gen. Rel. Grav.} \textbf{17} 343
\bibitem{fkn} Fritelli S, Kozameh C and Newman E T 1995 \emph{J. Math.
Phys.} \textbf{36} 4984
\bibitem{fknn} Fritelli S, Kozameh C, Newman E T and Nurowski P 2002 \emph{Preprint}
 http://www.fuw.edu.pl/\~{}nurowski/homepage.html
\bibitem{f} Friedrich H 1977 \emph{GRG Journal} \textbf{8} 303
\bibitem{chernmosser} Chern S S and Moser J 1974 \emph{Acta Math.} \textbf{133} 219
\bibitem{cartan} Cartan E 1932 \emph{Ann. Math. Pura Appl.} (4) \textbf{11} 17
\bibitem{f2} Friedrich H 1995 \emph{J. Geom. Phys} \textbf{17} 125--184
\bibitem{frauensp} Frauendiener J and Sparling G A J 2002 \emph{Preprint} arXiv:gr-qc/9907067
\end{thebibliography}
\end{document}